 \pdfoutput=1

\documentclass{jfm}

\usepackage{graphicx}
\usepackage{natbib}

\ifCUPmtlplainloaded \else
  \checkfont{eurm10}
  \iffontfound
    \IfFileExists{upmath.sty}
      {\typeout{^^JFound AMS Euler Roman fonts on the system,
                   using the 'upmath' package.^^J}%
       \usepackage{upmath}}
      {\typeout{^^JFound AMS Euler Roman fonts on the system, but you
                   dont seem to have the}%
       \typeout{'upmath' package installed. JFM.cls can take advantage
                 of these fonts,^^Jif you use 'upmath' package.^^J}%
      }
  \else
  \fi
\fi

\ifCUPmtlplainloaded \else
  \checkfont{msam10}
  \iffontfound
    \IfFileExists{amssymb.sty}
      {\typeout{^^JFound AMS Symbol fonts on the system, using the
                'amssymb' package.^^J}%
       \usepackage{amssymb}%
         
       \let\ge=\geqslant  
      }{}
  \fi
\fi

\ifCUPmtlplainloaded \else
  \IfFileExists{amsbsy.sty}
    {\typeout{^^JFound the 'amsbsy' package on the system, using it.^^J}%
     \usepackage{amsbsy}}
    {}
\fi

\newsavebox{\astrutbox}
\sbox{\astrutbox}{\rule[-5pt]{0pt}{20pt}}

\title[Turbulence cascade]{Fossil turbulence and fossil turbulence waves can be dangerous}

\author[C. H. Gibson]%
{C\ls A\ls R\ls L\ns H.\ns G\ls I\ls B\ls S\ls O\ls N$^1$%
  \thanks{Email address for correspondence: cgibson@ucsd.edu}}

\affiliation{$^1$Departments of MAE and SIO, Center for Astrophysics and Space Sciences, University of California at San Diego,
La Jolla, CA 92093-0411, USA\\}

\pubyear{2010}
\volume{650}
\pagerange{119--126}
\date{?; revised ?; accepted ?. - To be entered by editorial office}
\begin{document}

\maketitle

\begin{abstract}
Turbulence is  defined as an eddy-like state of fluid motion where the inertial-vortex forces  of the eddies are larger than any other forces that tend to damp the eddies out.  By this definition, turbulence always cascades from small scales where vorticity is created to larger scales where turbulence fossilizes.  Fossil turbulence is any perturbation in a hydrophysical field produced by turbulence that persists after the fluid is no longer turbulent at the scale of the perturbation.  Fossil turbulence patterns and fossil turbulence waves preserve and propagate energy and information about previous turbulence.  Ignorance of fossil turbulence properties can be dangerous.  Examples include the Osama bin Laden helicopter crash and the Air France 447 Airbus crash, both unfairly blamed on the pilots. Observations support the proposed definitions, and suggest even direct numerical simulations of turbulence require caution.
\end{abstract}

\begin{keywords}
Authors should not enter keywords on the manuscript, as these must be chosen by the author during the online submission process and will then be added during the typesetting process (see http://journals.cambridge.org/data/\linebreak[3]relatedlink/jfm-\linebreak[3]keywords.pdf for the full list)
\end{keywords}

\section{Introduction}

Turbulence is notoriously difficult to define.  Most discussions avoid explicit definition.  Instead, turbulence is identified intuitively from broad lists of known symptoms, like a disease (Tennekes and Lumley).  Here a narrow definition is required based on the inertial-vortex force $ \vec{v} \times \vec{w}$, where $\vec{v}$ is velocity and $\vec{w}$ is vorticity,  \cite{gib96}.  Flows not dominated by inertial-vortex forces are non-turbulent by this definition, \cite{gib86}.  Without this definition, one cannot distinguish between active turbulence and fossil turbulence flows, which have many important differences.  Persistent perturbations of vorticity, temperature, density etc. produced by turbulence at length scales of the fluid that are no longer actively turbulent are termed fossil turbulence, \cite{gib80}.  Because vorticity is produced at small viscous scales, turbulence must always cascade from small scales to large, contrary to claims of  \cite{tay38}, \cite{lum92} and the L. F. Richardson (1922) poem\begin{quotation}
Big whorls have little whorls, which feed on their velocity, and little whorls have lessor whorls, and so on to viscosity (in the molecular sense).
\end{quotation}  Irrotational flows are (by our definition) non-turbulent, even though irrotational flows typically provide the kinetic energy of the smaller scale turbulence.  Fossil turbulence and fossil turbulence waves preserve and propagate information about the original turbulence events, and generally dominate mixing and diffusion processes in natural fluids and flows, \cite{gib10}, \cite{gs10a}, \cite{gs10b}, \cite{gib12}, particularly in astrophysics, oceanography, atmospheric sciences, and cosmology.   

By making a distinction between turbulence and fossil turbulence, Kolmogorovian universal similarity laws of turbulence and turbulent mixing are provided a physical basis.  Kolmogorov scale vortex sheets are unstable to small scale vortex mergers driven
by inertial-vortex forces, demonstrating the universal turbulent cascade mechanism from small scales to large. 
With the present definition of turbulence, it is possible to unambiguously distinguish turbulent fluid from
non-turbulent fluid using hydrodynamic phase diagrams (HPDs), \cite{gib86}.  Because turbulence in natural fluids like the ocean
can be extremely intermittent, tens of thousands of HPDs were required to distinguish between turbulence and fossil
turbulence mixing mechanisms in a field test of submerged stratified turbulence using a municipal outfall, \cite{gbkl11}, \cite{leung11}.  A generic stratified turbulent transport mechanism was discovered termed Beamed Zombie Turbulence Maser Action Mixing
Chimneys (BZTMAMC) similar to that of electricity transport in lightning storms.  Heat, mass, momentum, and information
transport establish non-linear fossil turbulence internal wave vertical mixing channels similar to the ionized channels of repeated lightning strikes.  Maser action beaming in astrophysics is well known, \cite{alc86}.

In the natural flows of the ocean, atmosphere, and cosmology, many examples of fossil turbulence are found.  Jet aircraft wingtip vortices and contrails are fossils of turbulence that persist hours after all turbulence from the airplane has ceased at length scales of the contrails.  An important property of fossil vorticity turbulence is that most of the kinetic energy of the turbulence is preserved in its fossils and radiated by its fossil vorticity turbulence waves.  For this reason, small aircraft should not follow large aircraft closely.  Gliders should avoid mountains. Airlines crossing the equator during hurricane season should avoid thunderstorm regions.

  Fossils of big bang turbulence created at $10^{-35}$ m Planck scales have persisted for $\sim 13.7$ billion years at length scales $\ge 10^{25}$ m, \cite{gib04, gib05}. The spin direction of the big bang is preserved by that of the Galaxy and solar system, \cite{sg07}.   Direct numerical simulations of stratified turbulent wakes have demonstrated fossil turbulence wave radiation and highly persistent perturbations at Ozmidov scales $[L_R = (\varepsilon / N^3)^{1/3}]_0 $ of the turbulence at fossilization, \cite{pham09, brucker10, diam05, diam11}.  The viscous dissipation rate per unit mass is $\varepsilon$.  The stratification frequency is $N$.   

Neglect or ignorance of fossil turbulence and its properties can be dangerous.   The first helicopter carrying Seal Team Six to the Osama bin Ladin compound crashed for unknown reasons, despite numerous practice runs hovering safely over a full scale mock-up, \cite{owen12}.  Pilot error is assumed, but unanticipated amounts of dense night air seems more likely.  Air France 447 crashed crossing the equator on the first day of hurricane season, flying through a thunderstorm region.  Everyone was killed, and the pilots were blamed. 
 Icing and clear air turbulence guidelines to pilots have no latitude dependence, despite evidence
of maximum turbulence intermittency at the equator, \cite{baker87}.   
The President of Harvard University (1971-1991)  Derek Bok put it this way\begin{quotation}
If you think education is expensive, try ignorance.
\end{quotation}

A situation of particularly high danger would be ignorance of 
hostile submarines operating near our coasts.  
Consequences of this hazard is avoided by taking the 
properties of fossil turbulence and fossil turbulence waves into
account.  Most of the kinetic energy of turbulence from submarines in a stratified flow is transfered
with minimal loss to the kinetic energy of fossil turbulence 
patches (equations 2 below), which then radiate and reradiate
 the energy vertically as fossil and zombie turbulence waves to the surface where it may be 
detected (by BZTMAMC methods), along with persistent and detailed information about the turbulence sources.
Direct numerical simulations of stratified turbulent wakes have now confirmed persistence
for time periods $t \ge 1000 N^{-1}$, \cite{diam11}, confirming fossilization of the turbulence.
    
\begin{figure}
  \centerline{\includegraphics{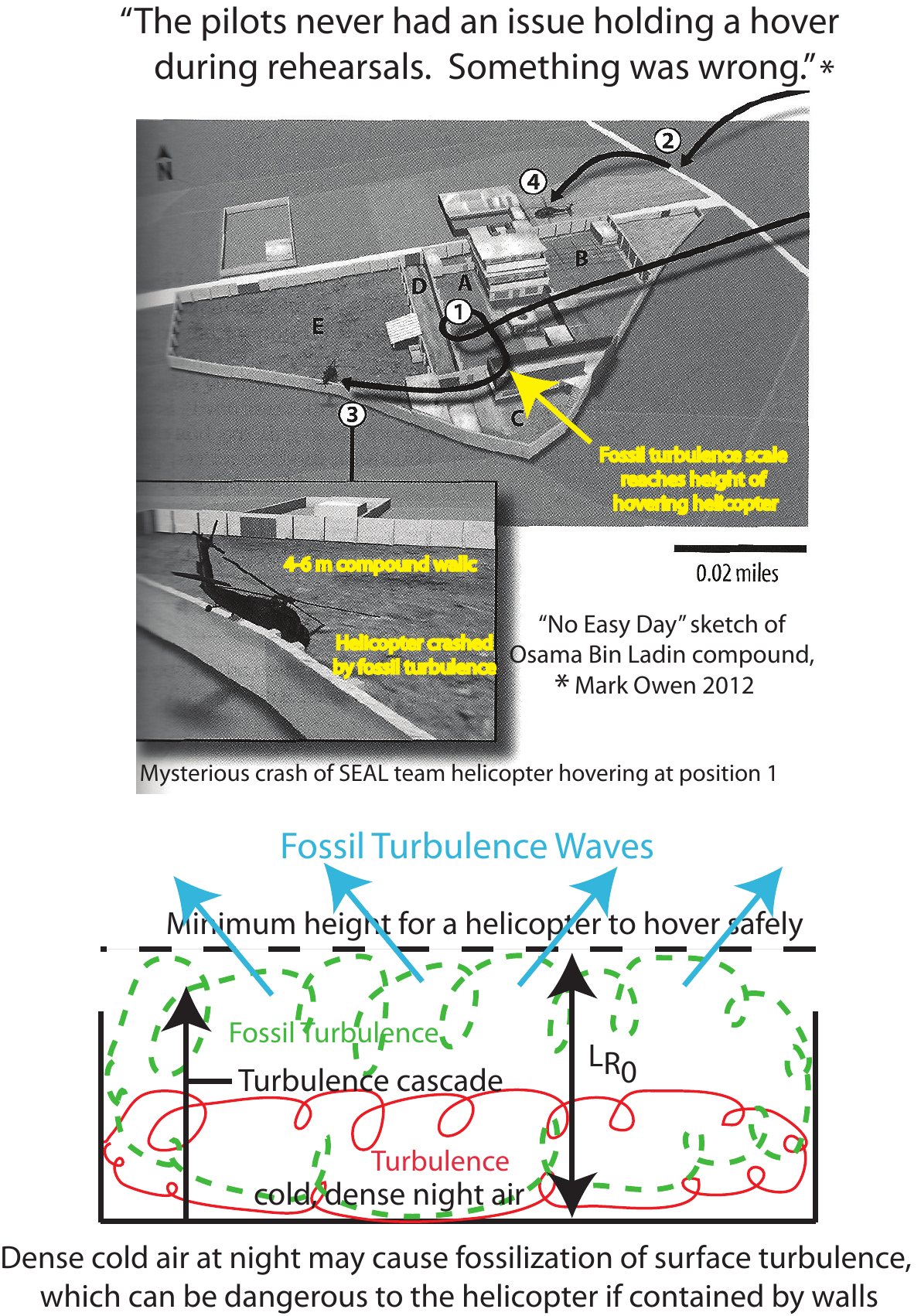}}
  \caption{Normally the downwash of a hovering helicopter is blown away and fossilizes elsewhere.  However, tall (4-6 m)
  walls apparently retained the dense fossil turbulence layer in the Osama bin Ladin compound.  
  Cold surface layer conditions on the clear night apparently  permitted growth  of the fossil
  turbulence thickness $\approx [L_R]_0$ to the minimum safe 
  hovering height of the helicopter (dashed line) $\approx 8$ m.}
\end{figure}

\section{Theory}
The conservation of momentum equations for collisional fluids may be written with the inertial-vortex force separated from the Bernoulli group of energy terms $B={v^2}/2 + p/{\rho + lw}$ and the various other forces of the flow.

\begin{equation}
  \partial{\vec{v}/\partial{t}}=-\nabla{B}+\vec{v} \times \vec{w}+\vec{{F_{viscous}}}+\vec{{F_{buoyancy}}}+\vec{{F_{Coriolis}}}+\vec{{F_{etc.}}}
  \label{Helm}
\end{equation}

For many natural flows the sum of the kinetic energy per unit mass ${v^2}/2$ and the stagnation enthalpy  $p/{\rho}$ along a streamline are nearly constant and the lost work $lw$ is negligible.  Turbulence is the class of fluid motions that arise when the first force term on the right $-\nabla{B} \sim 0$ and the the inertial vortex force $\vec{v} \times \vec{w}$ dominates all the other forces.

The ratio of the inertial vortex forces to the viscous forces of a a flow

\begin{equation}
Re = \vec{v} \times \vec{w} / \vec{{F_{viscous}}}
  \label{Helm}
\end{equation}

is the Reynolds number Re.  The best known criterion for the existence of turbulence is that $Re \ge Re_{crit}$, where $Re_{crit}$ is a universal critical value $\sim 10-100$ from the first Kolmogorov hypothesis. 

For turbulence to exist, inertial vortex forces must also overcome gravitational forces.  A Froude number $Fr$ ratio

\begin{equation}
Fr = \vec{v} \times \vec{w} / \vec{{F_{buoyancy}}}
  \label{Helm}
\end{equation}

 must exceed a universal critical value $Fr_{crit}$.  Many other dimensionless groups based on ratios of the inertial vortex force to other forces have been discussed.
 
We see that turbulence always must begin at the Kolmogorov length scale\cite{kol41} where the Reynolds number first exceeds a critical value and vorticity is produced by viscous forces.  The turbulence cascades to larger scales by vortex pairing until limited by one of the other fluid forces.  The Ozmidov scale at beginning of fossilization determines the maximum vertical scale of turbulence overturns and the scale of fossil turbulence waves radiated.  In self gravitational flows the buoyancy period is replaced by the gravitational free fall time. 

The proposed cascade of turbulent kinetic energy from small scales to large is the inverse of the  \cite{tay38} and \cite{lum92} cascades, both of which are physically backwards and misleading as seen by the growth of wakes, jets, and boundary layers.
According to the Lumley cascade theory, small scale turbulence does not appear for two overturn times $l/u$, where $l$ and $u$ are energy (Obukhov) scales of the turbulence.  This is illustrated by the Lumley spectral flux model shown in Figure 2, which is a non-turbulent energy cascade.  When turbulence is defined by the inertial vortex force, the cascade direction is always opposite to that proposed
by Taylor and Lumley.

\begin{figure}
  \centerline{\includegraphics{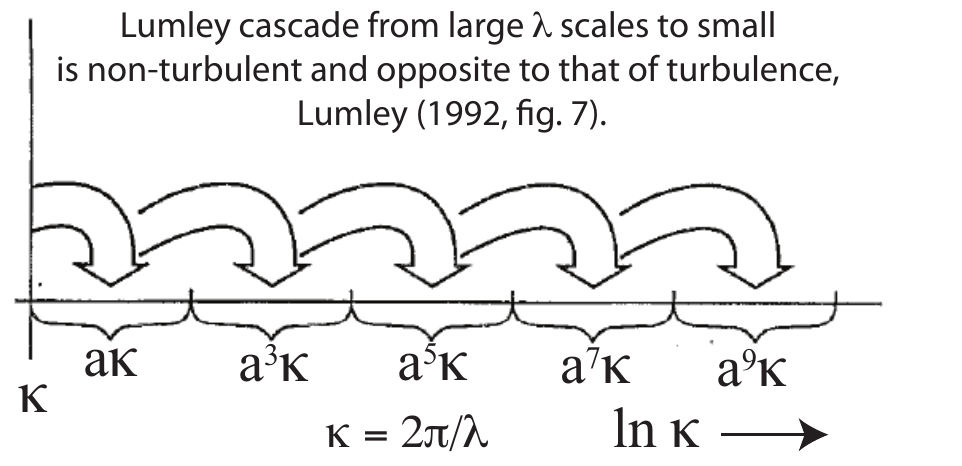}}
  \caption{Spectral flux model of Lumley (1992), suggesting small scale turbulence appears only after large scale turbulence
  has had time to cascade to dissipation scales.  The kinetic energy cascade shown is not a turbulence cascade because 
  the flow is irrotational.}
\end{figure}

\section{Observations}
\subsection{Laboratory}

The first and only laboratory evidence of fossil turbulence wave mixing (known by the author) is shown in Figure 3.  Highly concentrated
sodium nitrate solution is injected through downstream holes in a plastic cylinder.  No salt
should escape the turbulent portion of the cylinder wake.  Salinity spikes were detected
by a single electrode conductivity probe (1/4 micron) on the test section axis, indicating an unexpected salinity transport
mechanism must exist across the turbulence superlayer.  Baroclinic torques are maximum
at the top and bottom of the wake, as shown by the cartoon at the right of Fig. 3. Vorticity and turbulence
are produced, expanding the superlayer by turbulent mixing  on the bottom of the cylinder wake,
 but penetrating into the irrotational flow at the
top (dashed arrow) by fossil turbulence wave mixing.

\begin{figure}
  \centerline{\includegraphics{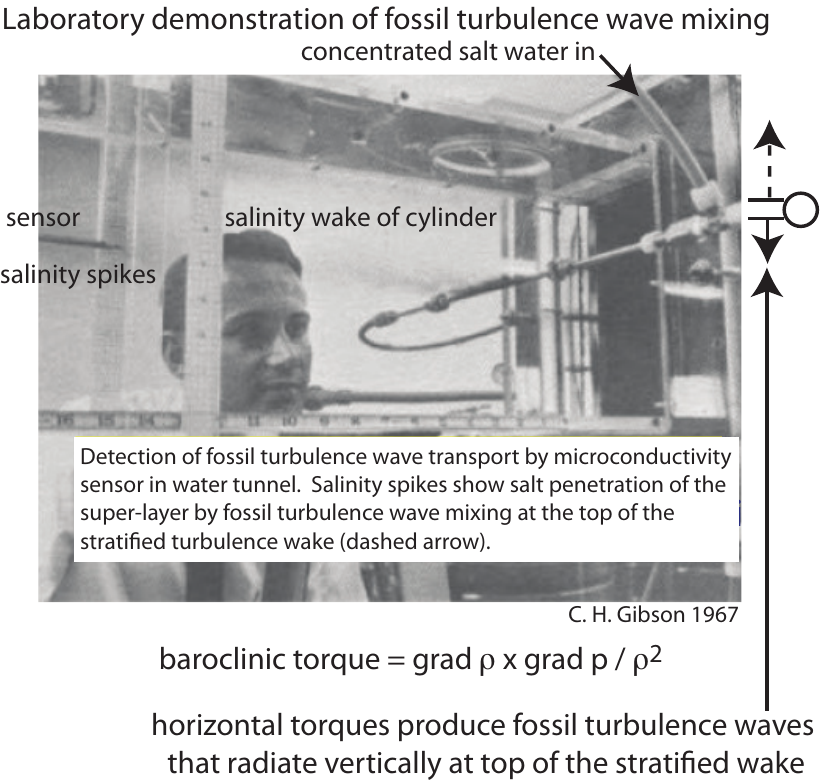}}
  \caption{Laboratory detection of fossil turbulence wave mixing, 
  Gibson 1967 AMES Department archives.  Natural flows such as the ocean, atmosphere, stars
  and the big bang depend on fossil turbulence wave mixing for transport of heat, mass, momentum
  and information to distances far from the turbulence source in non-turbulent fluid.}
\end{figure}

The wake superlayer (Stan Corrsin) is strongly distorted.  Unmixed irrotational
 fluid appears on the wake axis about 5\% of the time and
is easily discriminated from the turbulent fluid by its lack of salinity microstructure.
Rare conductivity spikes detected show fossil turbulence wave radiation has penetrated the turbulence superlayer, driven by the large baroclinic torques $\nabla \rho \times \nabla p / \rho ^2$ upstream (dashed arrow).  The Taylor-Lumley cascades of turbulence and turbulent kinetic energy from large scales to small, \cite{tay38, lum92}, are falsified by these observations showing the actual direction of the turbulence cascade is from small scales to large. 

In natural flows such as the ocean and atmosphere, turbulence is rare.  Most of the mixing is accomplished by fossil turbulence
and fossil turbulence waves.  Even the fossil turbulence is rare and intermittent, leading 
to the term dark mixing (Tom Dillon); that is, mixing
that must be there but is undetectable in most experiments that measure mixing.

\subsection{Cosmology}

According to hydrogravitational dynamics HGD cosmology (Gibson 1996, 2004, 2005) the universe began due to a turbulence instability at Planck conditions.  Evidence for a turbulent big bang event is emerging from space telescope observations of the cosmic microwave background CMB, \cite{gib10}.  The most recent evidence from WMAP CMB is in Figure 4, \cite{stark12}.  Starkman et al. emphasize
strong departures of the largest scale temperature-temperature (TT) correlations with the standard $\Lambda CDMHC$
cosmological model.  The departures are easy to understand using the proposed definition of turbulence, which cascades
from small scales to large and then fossilizes.  Fossil big bang vorticity explains the observation that the largest scales
observed in the CMB are aligned with a particular 
{``axis of evil``}
 direction of spin, contrary to isotropy assumptions of all cosmologies not driven by turbulence.

\begin{figure}
  \centerline{\includegraphics{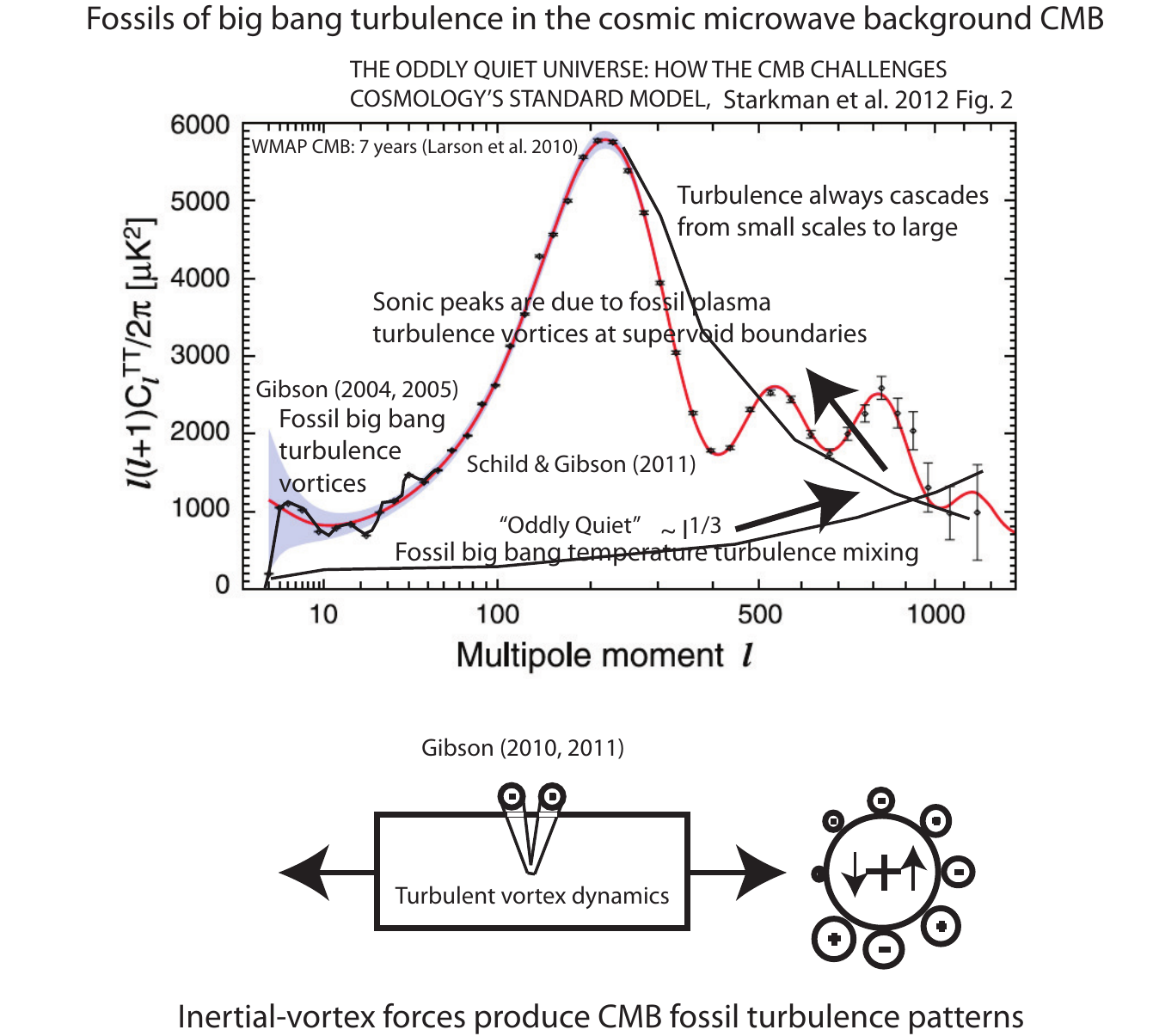}}
  \caption{Patterns that appear in cosmic microwave background $CMB$ spectra (top), Starkman et al. 2012,
  can best be described as fossils of turbulence that cascades from small scales to large, starting at Planck
  scales and described as the first turbulent combustion, Gibson (2004, 2005).  Spin  
  vectors at the largest scales are aligned (bottom cartoon) and have a characteristic pattern with secondary
  vortex peaks.
  Patterns observed in the CMB at the largest scales (grey shading) support turbulence as the source of the big bang event. 
  A similar pattern with vortex peaks is attributed to weak turbulence produced by viscous-gravitational fragmentation of
  the expanding primordial plasma to produce supervoids, Gibson (2010, 2011). }
\end{figure}

As shown in Fig. 4 (top), the measured spherical harmonics spectrum $C_\ell$ challenges the standard cosmological model, but is easy to understand as a fossil of big bang turbulence at multipole spatial frequencies $\ell$ near 10.    Multiple spectral peaks near $\ell = 10$ reflect vortex dynamics fossils of the big bang turbulence fireball, as shown by the stretching vortex
and secondary vortices in Fig. 4 (bottom), \cite{gib10,gsw11}. A similar fossil of turbulent supervoid fragmentation of the plasma along vortex lines explains $C_{\ell}$ for $\ell$ values larger than 200.  Baryon oscillations in cold dark matter potential wells are not needed to account for the sonic peaks observed at $\ell \ge 200$ and would be damped by the large photon viscosity of the plasma, even if cold dark matter CDM potential wells were not mythical. 

Vortex line stretching provides the anti-gravitational negative stresses needed to extract mass-energy from the vacuum, as shown (arrows).  Dark energy $\Lambda$ is not needed.  Fossilization of big bang turbulence, \cite{gib05}, is due to phase changes of the Planck fluid as temperatures decrease from $10^{32}$ K to $\sim 10^{28}$ K, so that quarks and gluons, and large gluon viscosity, become possible, \cite{gib04}.  

Small scale turbulence during the plasma epoch was prevented by photon viscous forces from inelastic scattering of photons by electrons until the Schwarz viscous-gravitational scale $L_{SV}$ matched the scale of causal connection $L_H = ct$ at time $t=10^{12}$ s, where $c$ is the speed of light, \cite{gib96}.  Plasma fragmentation fossils persist as the largest objects, from superclusters to galaxies, \cite{gs10a, gs10b}.   The Corrsin-Obukhov fossil temperature turbulence mixing subrange $\ell(\ell + 1)C_{\ell} \sim \ell ^{1/3}$ at the bottom of Fig. 4 is a cascade from large scales to small (lower arrow) produced by big bang turbulent mixing.  The $sonic$ $peaks$ at $\ell \ge 200$ reflect the turbulence cascade (upper arrow) from small scales to large at the boundaries of gravitationally expanding super-cluster-voids.  The speed of expansion of voids is limited by the speed of sound $c/3^{1/2}$ in the plasma, and $\ell =200$ reflects the time  between first fragmentation time $t \sim 10^{12}$ s (30 000 years) and when the plasma becomes gas at $t \sim 10^{13}$ s.  

Because the kinematic viscosity $\nu$  decreases by a factor of $10^{13}$ from that of the plasma, the gas then fragments into planets in dense clumps that may become globular star clusters, \cite{gib96}.  This was inferred independently by \cite{schild96}  from his careful observations of gravitational lensing of a distant quasar by an intervening galaxy over a fifteen year period.  Because the twinkling period of quasar images is weeks (planets) rather than years (stars), Schild could infer that the mass of the galaxy is dominated by planets rather than stars.  Differences in brightness of the quasar images show the planets are in clumps.  
  
\subsection{Stratified shear flows and wakes}

As we have seen, the definitions of turbulence and fossil turbulence are important to understanding cosmology.  They are equally important to understanding terrestrial flows. High resolution microstructure detectors reveal most of oceanic mixing takes place in the fossil turbulence state, \cite{leung11}.  A generic mechanism of turbulent-fossil turbulence wave energy and information  transport from small scales to large is indicated, \cite{gbkl11}.  To capture the full stratified turbulence cascade mechanism, direct numerical simulations are required that resolve the Kolmogorov length scales where vorticity and turbulence are generated, \cite{diam05}, with a range of stratification periods ($\ge 1000$) sufficient to unequivocally demonstrate fossilization
and with allowance for the evolution of the flow, \cite{diam11}. Numerical simulations of turbulence require caution if energy is inserted at large scales rather than small, and if time evolution of the flow is not included.

 \begin{figure}
  \centerline{\includegraphics{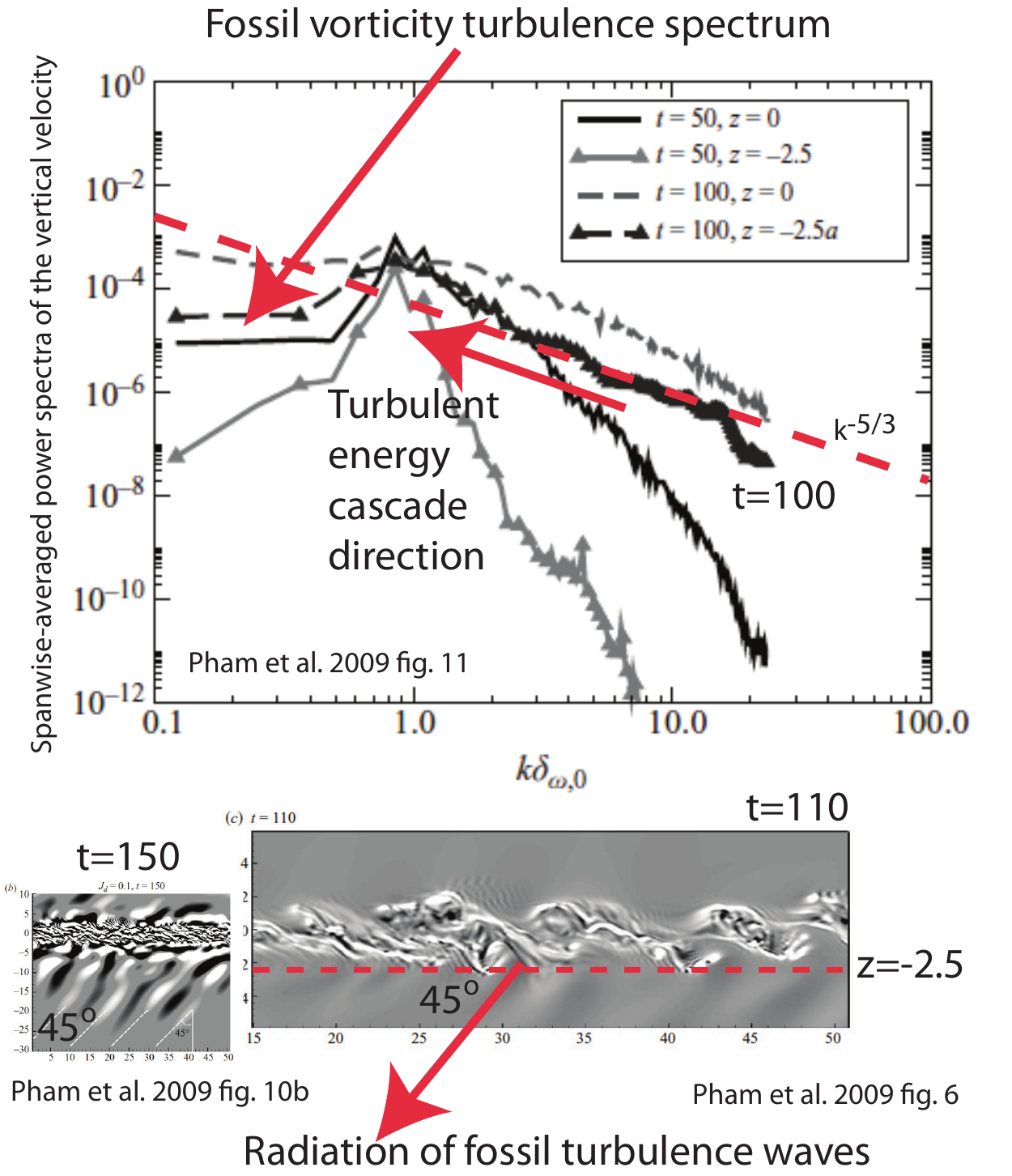}}
  \caption{Direct numerical simulation of $x,y,z,t$ turbulence and fossil turbulence waves formed by shear, \cite{pham09}.}
\label{fig:kd}
\end{figure}
 
 A fossil vorticity turbulence spectrum ($t=100, z= -2.5$) has been captured using direct numerical simulations of sheared stratified turbulence, as shown in Figure 5, from Fig. 11 in \cite{pham09}.  The dashed line with slope -5/3 suggests turbulence at small scales cascades to the Ozmidov scale at fossilization, the thickness of the shear layer.  Radiation of fossil turbulence waves at the Ozmidov scale and expected angle $45^o$ is shown at the bottom of Fig. 5.
 
Direct numerical simulations of stratified turbulent wakes are compared to non-stratified wakes by \cite{brucker10}, Figure 6.
As shown in Fig. 6 (top), turbulence forms in both stratified and non-stratified turbulent wakes when the Obukhov energy scale $L_O$ exceeds the Kolmogorov scale $L_K$.  The non-stratified turbulent wake cascades to larger length scales and vanishes without a trace.  The stratified turbulent wake fossilizes when the Ozmidov scale becomes smaller than the Ozmidov scale at beginning of fossilization.  Fossil turbulence waves propagate into surrounding stratified layers and persist for periods of time exceeding 1000 stratification periods (bottom), confirming the Diamessis et al. (2011) computations.

\begin{figure}
  \centerline{\includegraphics{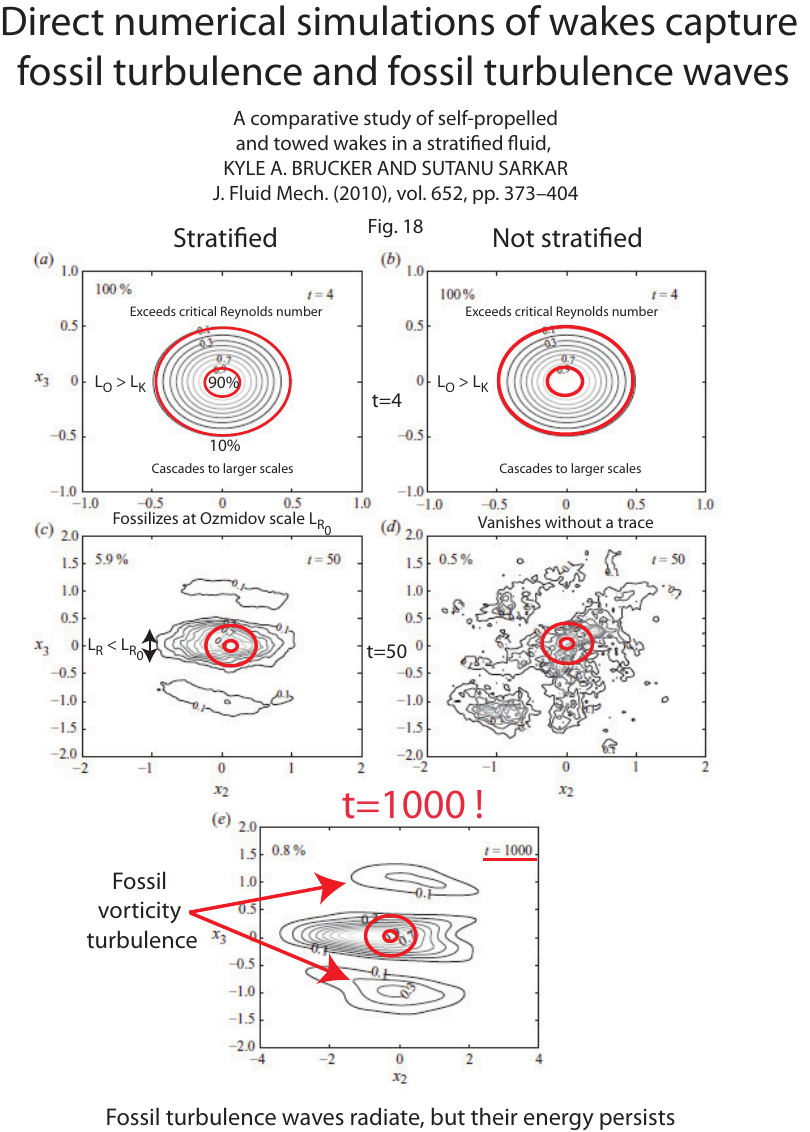}}
  \caption{Direct numerical simulation of $x,y,z,t$ turbulence and fossil turbulence waves formed by stratified and non-stratified wakes, Brucker et al. 2010.}
\label{fig:kd}
\end{figure}

\subsection{Crash of Air France 447}

An important property of stratified and rotating turbulence is the intermittency of the viscous dissipation rate of turbulence, described
by the Kolmogorov third universal similarity hypothesis.  Nonlinear processes such as gravity, personal income, self-gravitational density fluctuations and biological activity
are examples of other physical processes that are statistically similar to turbulence.  Such random variables $x$ are generically
lognormal and may be extremely intermittent.  Intermittent random variables are subject to large undersampling errors.  The intermittency factor of $x$ is defined as the variance of the natural logarithm ${\sigma ^2}_{ln x}$.  On the Equator, the intermittency 
factors for turbulence and mixing are estimated to be about 7, \cite{baker87}, causing the probable undersampling errors
to exceed $10^4$.  This is the mean to mode ratio of a lognormal distribution with intermittency factor 7.  At midlatitudes the intermittency factor for $\varepsilon$ is about 5, so the mean to mode ratio is about 2000.

\begin{figure}
  \centerline{\includegraphics{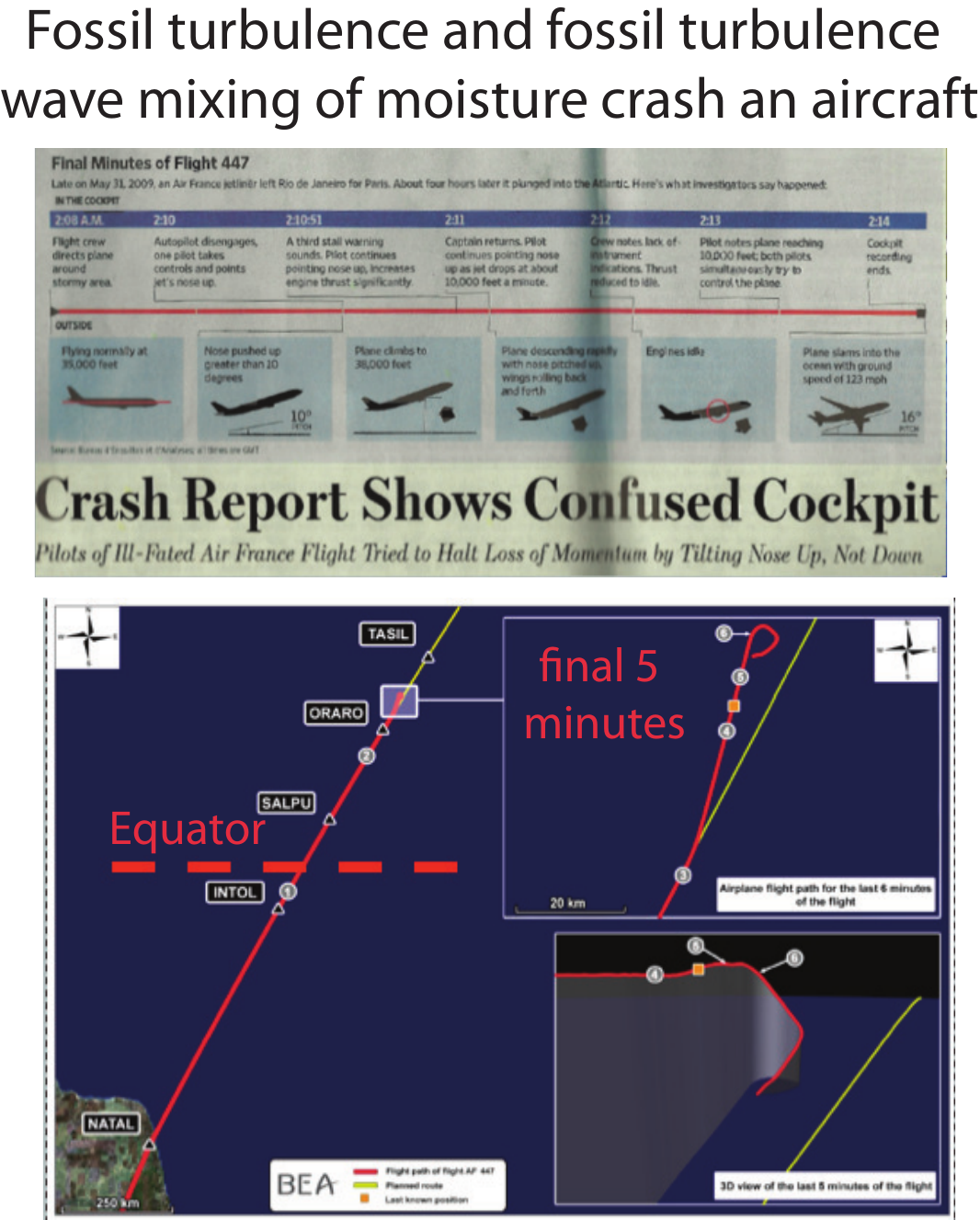}}
  \caption{Air France 447 crash is attributed to a combination of highly intermittent equatorial turbulence and 
  powerful thunderstorm mixing of moisture by fossil turbulence waves.  Ignorance of these factors permitted pilots to cross the Equator on the first day of hurricane season through a thunderstorm region.  The Wall Street Journal report (top) shows the plane
  vastly lost lift.  The BEA report (bottom) shows the plane also vastly lost control of direction.  If the icing mixing layer
  thickness $[L_{R_{0}}]_{icing}$ in the thunderstorm exceeded the altitude, the plane was doomed independent of pilot actions.}
\end{figure}

\begin{figure}
  \centerline{\includegraphics{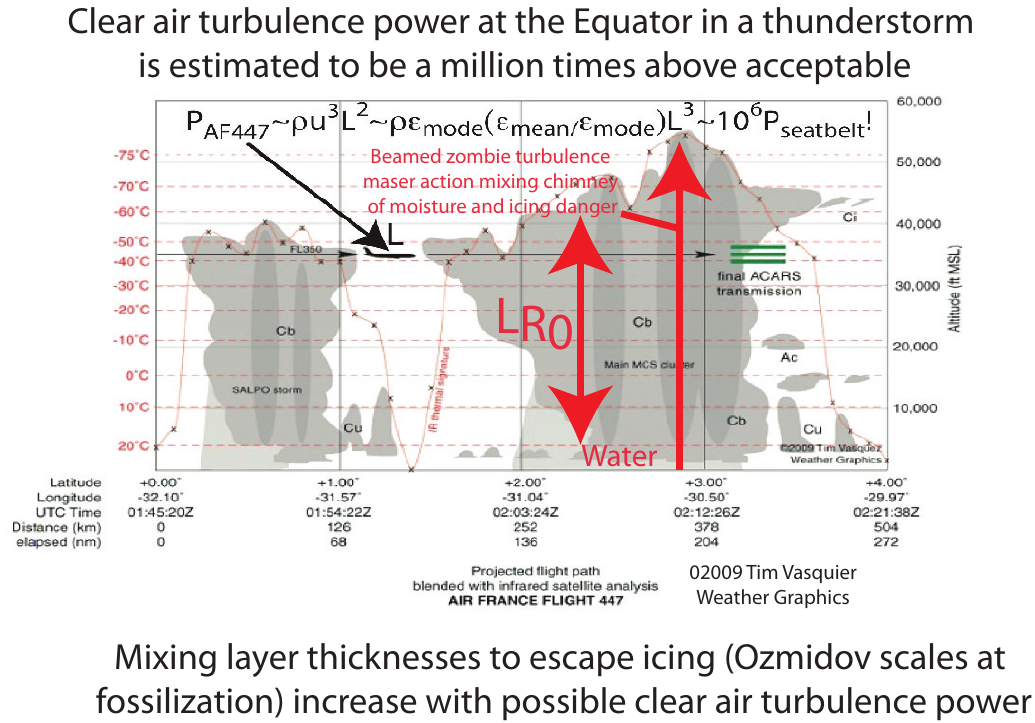}}
  \caption{The Air France 447 flight path is shown, terminating in the water.}
\end{figure}

Figure 7 shows the final stages of the Air France 447 flight across the equator on May 31 to June 1 2009, the first day of hurricane season.  The newspaper account (Fig. 7 top) showed the pilots first attempted to gain altitude when they encountered evidence of icing, but soon lost control of lift and direction.

The BEA (Bureau d'Enqutes et d'Analyses pour la sŽcuritŽ de l'aviation civile) report of the final 5 minutes of the flight is shown
at the bottom of Fig. 7.  

  Figure 8 shows the flight path through powerful thunderstorm regions.  An expression is given estimating the maximum power
  of a clear air turbulence event at the Equator, taking intermittency into account.   Assuming the viscous dissipation rate of 
  turbulence is a hundred times larger than average when the seat belt sign goes on suggests approximately a million times larger
  power might be exerted on the aircraft of scale L.  The evidence of Fig. 7 suggests it was not this power that resulted in the
  crash, but the resulting increased probability of icing.

\subsection{Oceanography and fossil turbulence}

Most of the mixing of temperature, salinity and vorticity fluctuations in the ocean is in fluid which has been scrambled by turbulence, but is no longer turbulent at the scale of the fluctuation; that is, it is fossil turbulence, by definition, \cite{gib80}. Oceanographers have
generally failed to recognize the profound differences between fossil and active turbulence processes.

\begin{figure}
  \centerline{\includegraphics{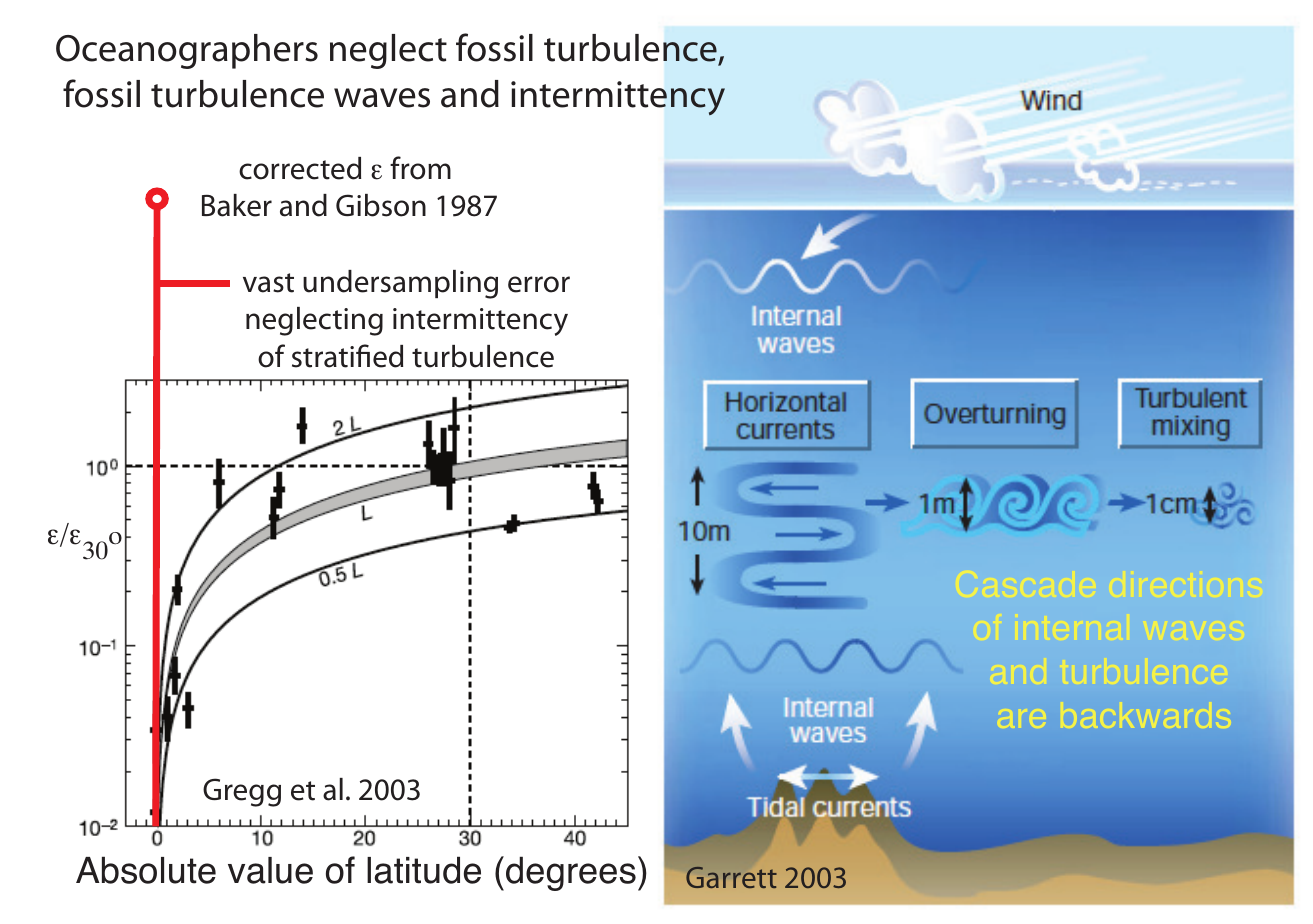}}
  \caption{Garrett and Gregg et al., Nature (2003).  Garrett (2003)  claims the Gregg et al.  data (left) 
  confirm his theoretical model (right), where large scale internal waves generated by wind at the surface
  and tides on the bottom cascade to small scale internal waves and currents that
  overturn and cause turbulence and mixing in the ocean interior.  These
  cascade directions are incorrect and backwards for internal waves, turbulence and turbulent mixing. }
\end{figure}

Figure 9 shows mean viscous dissipation rates $\varepsilon_{latitude} / \varepsilon _{{30}^{o}}$ estimated by oceanographers as a function of latitude, without taking fossil turbulence
and fossil turbulence waves into account, \cite{gregg03}, \cite{gar03}.  These mean values are vastly inaccurate, as a result
of misunderstandings of turbulence, fossil turbulence, fossil turbulence waves, and the directions of the internal wave and turbulence
cascades, and with no attempt to correct for the extreme intermittency of equatorial turbulence, \cite{baker87}.  

A red circle in Fig. 9 shows $\varepsilon_{equator} / \varepsilon _{{30}^{o}} \approx 10^{-2}$ multiplied by the mean to
mode ratio 30,000 expected by Baker and Gibson, corrected to give a value of 300.  Thus, the claim of small equatorial mixing fails
 due to large equatorial turbulence intermittency,
from both quantitative and qualitative undersampling errors.  As shown in Fig. 7 and Fig. 8, ignorance
of such critical properties of stratified, rotating,  turbulence can be dangerous.  Garrett praises the Gregg et al. 2003 results with
the statement,  \begin{quotation}
This cascade to smaller scales is reminiscent of what occurs in turbulent motions in unstratified water, with big eddies tearing one another apart and giving rise to ever smaller eddies, \end{quotation} 
which confirms the widely accepted Garrett theory of ocean mixing, based on the widely accepted (but equally questionable)
Taylor-Richardson-Lumley large-to-small-scale energy cascade theory of turbulence.

The \cite{gar03} concepts of internal waves and turbulent mixing shown in Fig. 9 oversimplify the underlying processes.  As we have seen, turbulence always starts from small scales and cascades to large, where fossilization processes begin.  Transport in the vertical direction is limited by buoyancy forces in the ocean and atmosphere, and by Coriolis forces in the horizontal direction.  Because the Coriolis forces at the equator go to zero, the cascade covers an enormous range of scales.  This leads to large intermittency factors that must be taken into account in estimating mean mixing and dissipation rates.  The actual mixing process is best described as a complex interaction between internal waves and turbulence termed Beamed Zombie Turbulence Maser Action mixing chimneys (BZTMA), \cite{gbkl11}.  This was studied carefully in a three year series of experiments testing satellite detected
 surface manifestations of a submerged municipal outfall in Hawaii, \cite{g07}, \cite{k05}.  
 
 It was found that
 large fossil turbulence patches produced by the outfall drifted into Mamala bay of Oahu, and would induce secondary (zombie)
 turbulent mixing and radiation to the surface by extracting energy from fossil turbulence internal waves radiated from the bottom. BZTMA mixing chimney processes dominated the mixing processes of the bay, and are suggested as generic to mixing processes in natural fluids such as the atmosphere, ocean, and in stars.
 
 Because turbulence must cascade from small scales to large because inertial vortex forces exist and always create the first turbulence in all flows at the Kolmogorov (inertial-viscous) scales, we see that attempts to carry out meaningful numerical 
 simulations of turbulence are limited in a fundamental way.  Only direct numerical simulations are reliable, and even these
 can be misleading if the time variation, intermittency and cascade direction of turbulence are ignored.  Large eddy
 simulations are clearly only as good as the empirical modeling of the small scale motions.  Reynolds Averaged Navier Stokes (RANS) equations may or may not capture large scale effects on the turbulence introduced at small scales.  This is demonstrated by a variety of seal species that can use their whiskers (vibrissae) to detect sophisticated properties of fossil vorticity turbulence wakes left by prey that  survive long after the turbulence has been damped by stratification, \cite{sp07}, \cite{mier11}.  Weddell seals, for example, have no need to migrate during the dark Antarctic winter, but are able to take advantage of the persistence and complex anisotropies of fossil vorticity turbulence to find food, and to detect and follow their own complex and persistent
 fossil vorticity turbulence wakes miles back to their air holes in the ice.
 
Field measurements of turbulence and mixing in the ocean and atmosphere that fail to include hydrodynamic phase diagrams taking into account BZTMA mixing chimney processes are also unlikely to produce reliable results.

An example of failed direct numerical simulations of turbulent mixing has recently been kindly provided by P.K. Yeung and K.A. Sreenivasan (personal communication of preprint submitted to JFM 10/6/2012).  In a direct numerical simulation of turbulent mixing
of a strongly diffusive scalar it was found that a -17/3 subrange (Batchelor Howells Townsend 1959) rather than -3 subrange (Gibson 1968ab) for the scalar spectrum was indicated.  This result is contrary to a variety of experiments involving tests of temperature spectra and other statistical parameters in turbulent mercury, as well as electron density measurements in atmospheric wakes, \cite{gib88}.

Figure 10 shows a comparison of scalar spectral forms measured by radio telescopes in the Galaxy. What is expected for the scalar is a -5/3 range followed by either a -17/3 or a -3 subrange.  What was found by the Yeung and Sreenivasan DNS simulation is -17/3 with no -5/3 and no -3.   In Fig. 10, one dimensional spectra are plotted, so -17/3 becomes -23/3, etc.  As shown, the Gibson -3 (-15/3)
spectral form is supported (red dashed line) by the observations, as well as the Corrsin-Obukhov -11/3 (-5/3) scalar inertial subrange, neither of which was detected by the Yeung-Sreenivasan DNS numerical simulation.  

Extremely high spatial and temporal resolution are required to capture the physical mechanisms of strongly diffusive turbulent mixing.
Uniform scalar gradients must be simulated at large scales.  Turbulent eddies must cascade from small scales to large and distort
these gradients to produce diffusive instabilities so zero scalar gradient objects are produced according to the Gibson 1968ab theory.
It is suggested that the simulation simply lacked adequate resolution to reproduce stratified turbulent mixing of strongly diffusive scalars.

\begin{figure}
  \centerline{\includegraphics{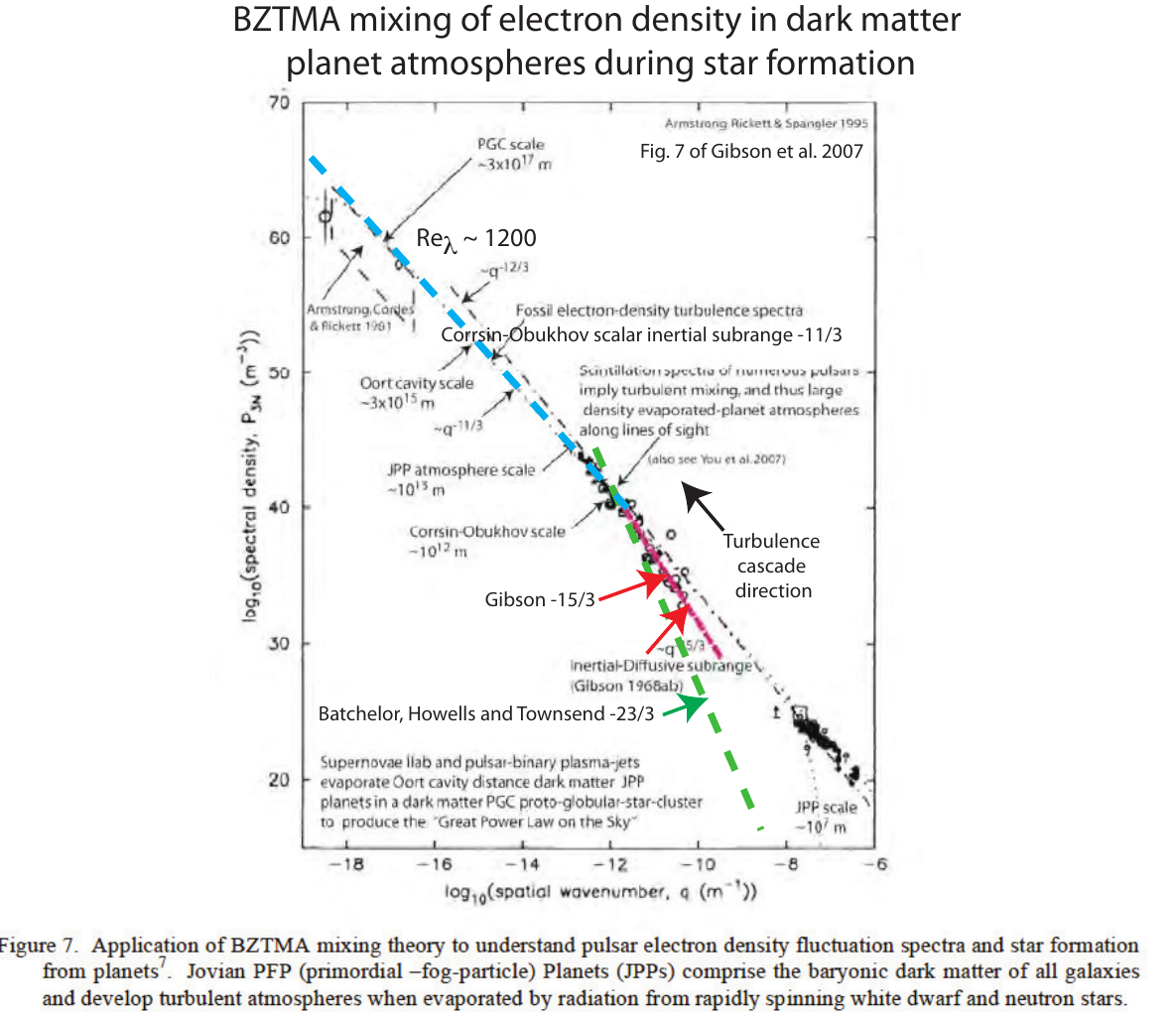}}
  \caption{BZTMA spectra of electron density in dark matter planet atmospheres show
  better agreement with the Gibson 1968ab spectral form (dashed red line)
  than with the Batchelor et al. 1959 prediction (dashed green line): Fig. 7 from \cite{g07}.  The -5/3
  Corrsin-Obukhov scalar-inertial subrange (dashed blue line) extends for six decades as part of the twelve decade
  Great Power Law on the Sky.  It is a fossil electron-density turbulence spectrum mixed by small scale turbulent
  jets from merging dark matter planets as they form Jupiters (JPPs) and stars in our proto-globular-star-cluster PGC clump
  of a trillion frozen gas planets, \cite{gib96}, \cite{schild96}, \cite{gs10b}.   }
\end{figure}

 BZTMA mixing of electron density accounts for the strong scattering of radio waves in the atmospheres of density stratified turbulent atmospheres of dark matter planets responsible for the formation of stars.  The dark matter planets form at the plasma to gas transition time (300,000 years) in clumps of a trillion termed proto-globular-star-clusters (PGCs).  All stars are formed in these PGC clumps.  The planets (mostly Earth mass) surrounding the stars are ionized by the new stars to provide large atmospheres of evaporated hydrogen and helium, and the electron density fluctuations that scatter the pulsar (spinning neutron star) radio waves detected in Fig. 10.   
 
BZTMA mixing of self-gravitational density with weak turbulence accounts for the remarkably uniform composition of the
gas emerging from the hot plasma epoch, producing photon-viscosity scale proto-galaxies uniformly fragmented into
PGC mass clumps of dark matter hydrogen-helium gas-viscosity planets.  Future studies of electron density spectra
can take full advantage of this uniformity and the universal similarity of turbulence and turbulent mixing, preserved
by the million PGCs of each galaxy, each with their trillion earth-mass planets merging to make stars, and the small
scale turbulence and fossil turbulence remnants that reveal their presence.

 \begin{figure}
  \centerline{\includegraphics{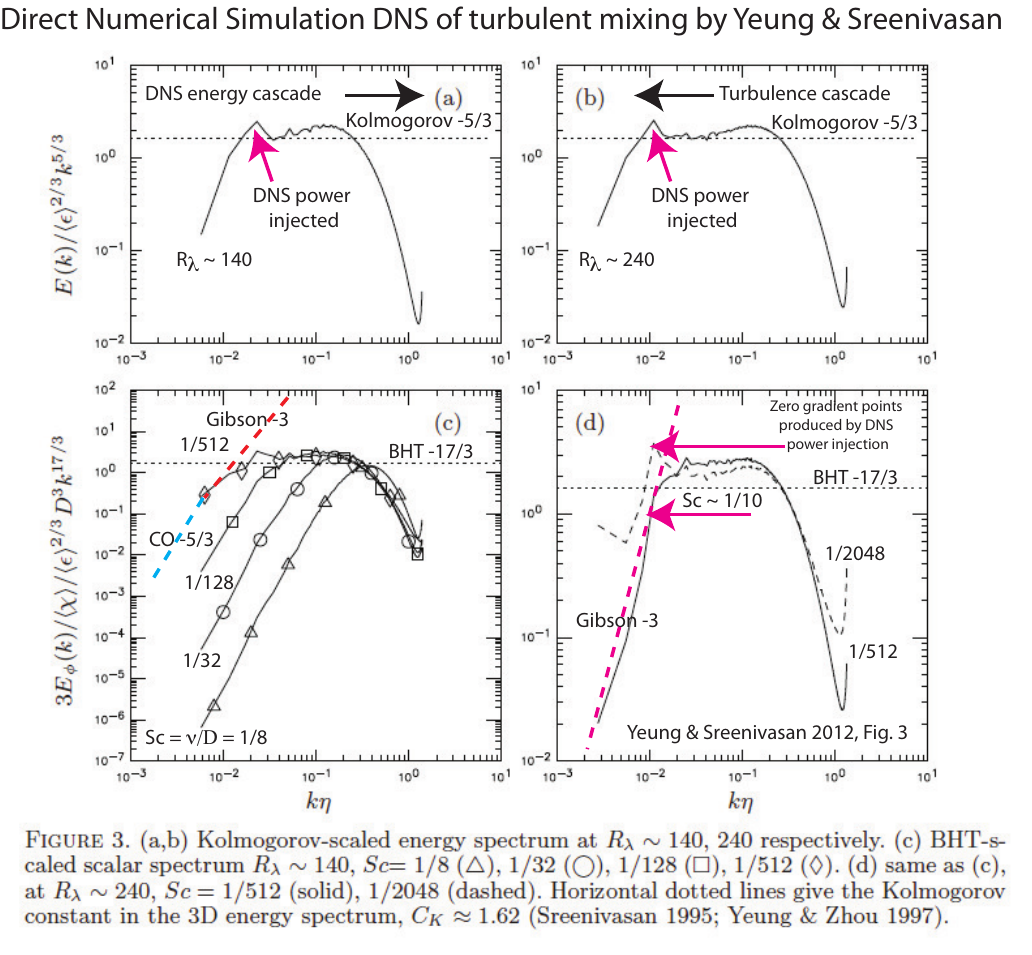}}
  \caption{DNS (Direct Numerical Simulation) of small Schmidt number turbulent mixing, \cite{ys12}, Fig. 3.  To properly
  simulate the physical mechanisms of such small Sc values (eg: zero scalar gradient point production) requires a
  wider range of length scales than presently available even on the most powerful computers, \cite{gib88}. As shown in Fig. 11d,
  the effective Schmidt number to produce zero gradient points and a -3 scalar subrange is Sc = 1/10 rather
  than 1/512.    }
\end{figure}

 Figure 11 shows a DNS numerical simulation of low Schmidt number turbulent mixing, \cite{ys12}, compared to spectral subrange
 predictions of Batchelor, Howells and Townsend, Gibson, and Corrsin and Obukhov tested by the radio telescope observation in Fig. 10, and temperature in mercury, \cite{gib88}.  Power to sustain turbulence is injected at small wavenumbers $k$, as shown by the peaks in the velocity and scalar spectra.  However, the resolution of the simulation only permits an effective Sc value of $\approx$ 1/10, as shown by the red arrows in Fig. 11d, lower right.  
 
 Taylor microscale Reynolds numbers  $R_{\lambda}$ of the numerical simulations were 140 and 240, well above the minimum required for turbulence to exist.  Universal dimensionless spectral forms for velocity (top) and scalar variance (bottom) are shown normalized by Kolmogorov variables.  As expected, a wider Kolmogorov -5/3 inertial subrange was found for the larger Reynolds number value, Fig. 11. top, ab.  Extrapolating the inertial subrange widths of Fig. 11 to the turbulence 
 of Fig. 10, with six decades of inertial subrange, gives  $(R_{\lambda})_0 \approx 1200$ for the turbulence produced
 by gas jets of the merging planets at the beginning of fossilization, which extend across the $3 \times 10^{17}$ m
 size of a PGC dark matter planet clump.
 
 \section{Discussion}
 
 Ignorance of a precise definition of turbulence based on the inertial vortex force prevents the possibility of recognizing
 fossil turbulence as a means of preserving information about previous turbulence, 
 and fossil turbulence waves as a means of propagating this
 information elsewhere.  Universal similarity laws of turbulence and turbulent mixing are vastly extended by fossil turbulence
 patterns and statistical parameters, especially in natural flows of the ocean, atmosphere, astrophysics and cosmology. 
 
Safety of aircraft is particularly sensitive to effects of turbulence, turbulent mixing, fossil turbulence
and fossil turbulence waves.  Two examples given of death and near-death show the heavy price that can 
result from ignorance of fossil turbulence effects.

National security interests are endangered by assumptions that submarines are invisible
by all means non-acoustical.  Clearly complex information about previous
turbulence persists in the wakes of fish and seals, \cite{mier11}, \cite{sp07}, long after their actively
turbulent patches have been damped by stratification.  It is not safe to assume that similar information
is not preserved in the wakes of submarines and propagated to the sea surface
and elsewhere by fossil turbulence waves.

\section{Summary and Conclusions}

Turbulence should be defined in terms of the inertial vortex forces of its eddy motions. It must be so defined for universal similarity laws of Kolmogorov, Batchelor etc. to apply.  Irrotational flows are recognized as non-turbulent by this definition.  Turbulence always cascades from small (Kolmogorov) scales to larger Ozmidov etc. scales where other forces or fluid phase changes cause it to fossilize.  Fossil turbulence preserves information about previous turbulence and continues the mixing and diffusion started by turbulence.  Fossil turbulence waves dominate the transport of heat, mass, chemical species and information in natural fluids and help preserve turbulence and mixing information in detectable forms.  The 1938 G. I. Taylor concept that turbulence cascades from large scales to small is incorrect and misleading, and should be abandoned.  Irrotational flows in the Lumley (1992) spectral pipeline model of the turbulent kinetic energy cascade from large scales to small should be identified as non-turbulent. 


\bibliographystyle{jfm}

\bibliography{jfm-instructions}

\end{document}